\def\bd{\begin{displaymath}}\def\ed{\end{displaymath}}
\def\be{\begin{equation}}\def\ee{\end{equation}}
\def\bea{\begin{eqnarray}}\def\eea{\end{eqnarray}}
\def\ba{\begin{array}}\def\ea{\end{array}}
\def\lb{\label}

\def\b{\beta}\def\c{\chi}

\def\k{\kappa}\def\l{\lambda}\def\m{\mu}\def\n{\nu}\def\o{\omega}\def
\p{\pi}\def\t{\tau}
\def\y{\eta}

\def\D{\Delta}

\def\inf{\infty}\def\id{\equiv}\def\mo{{-1}}

\def\mn{{\m\n}}

\def\eom{equations of motion }
\def\coo{coordinates }

\def\pb{Poisson brackets }\def\db{Dirac brackets }

\def\PL#1{Phys.\ Lett.\ {\bf#1}}
\def\PRL#1{Phys.\ Rev.\ Lett.\ {\bf#1}}
\def\PR#1{Phys.\ Rev.\ {\bf#1}}\def\CQG#1{Class.\ Quantum Grav.\ {\bf#1}}
\def\NP#1{Nucl.\ Phys.\ {\bf#1}}

 \def\IJMP#1{Int.\ J. Mod.\ Phys.\ {\bf #1}}
 
\def\PRep#1{Phys.\ Rep.\ {\bf#1}}

\def\JHEP#1{JHEP\ {\bf#1}}
\def\RMP#1{Rev.\ Mod.\ Phys.\ {\bf#1}}
\def\arx#1{{\tt arXiv:#1}}

\documentclass[12pt]{article}
\begin{document}
\def\J{J_{10}}\def\sd{\mathop{\rm sd}\nolimits}\def\cd{\mathop{\rm cd}}\def\nd{\,{\rm nd}}
\begin{titlepage}
\title{Classical dynamics on Snyder spacetime}
\bigskip
\author{ S. Mignemi\medskip\\ \small Dipartimento di Matematica e Informatica,\\
\small Universit\`a di Cagliari, viale Merello 92,\\\small 09123 Cagliari, Italy}
\maketitle

\begin{abstract}
We study the classical dynamics of a particle in Snyder spacetime, adopting
the formalism of constrained Hamiltonian systems introduced by Dirac.
We show that the motion of a particle in a scalar potential is deformed
with respect to special relativity by terms of order $\b E^2$. An important
result is that in the relativistic Snyder model a consistent choice of the
time variable must necessarily depend on the dynamics.

\end{abstract}

\end{titlepage}

\section{Introduction}
The interest on noncommutative geometries has greatly increased during last
years, because they may describe the structure
of spacetime at Planck scales, where the effects of quantum gravity are
sensible  and the location of particles in space and time may
become fuzzy \cite{ncg}.

Historically, the first example of noncommutative geometry was proposed by
Snyder \cite{Sn}, and was based on a deformation of the Heisenberg algebra
of quantum mechanics. In spite of the presence of a fundamental length scale,
his model is invariant under Lorentz transformations,
and only the action of the translations is not trivial \cite{BK,SM}.
The possibility of preserving the Lorentz invariance is due to the fact that
the deformed Heisenberg algebra is not a Lie algebra, as in simpler
noncommutative models \cite{ncg}, but rather a function algebra, the structure
constants being dependent on position and momentum.

Although the investigation of Snyder spacetime has been neglected for many
years, recently its implications have been studied from several points of
view \cite{Mi,others}.
In particular, in order to clarify the physical properties of the Snyder model,
it can be useful to start from the investigation of its classical limit.
This is described by a phase space with noncanonical symplectic structure, and
its dynamics must therefore necessarily be investigated using Hamiltonian methods.

In this framework, the classical motion of a nonrelativistic particle in Snyder
space has been studied in detail \cite{Mi}, and the exact solutions of the \eom
have been found in the case of a free particle and of a harmonic potential.
It results that, while the free motion is trivial, the classical dynamics is
modified in the presence of external forces. For example, the motion of a harmonic
oscillator is still periodic, but no longer given by a simple trigonometric
function as in classical mechanics, and the frequency of oscillation acquires
a dependence on the energy, like in special relativity.

It is interesting to investigate if these features extend to the relativistic
dynamics. The problem is not trivial, because it is known that in the relativistic
domain the Hamiltonian dynamics of a particle is constrained, and to treat the
problem one must employ for example the Dirac formalism \cite{RT,Mu}.
Moreover, due to the nontrivial \pb between time and
spatial coordinates of the relativistic Snyder model, its nonrelativistic limit
does not necessarily coincide with the nonrelativistic theory.

While the study of the motion of a free relativistic particle presents no problems
and reproduces the results of special relativity, the dynamics of a particle coupled
to an external potential of scalar type in Hamiltonian form is not well known
even in standard special relativity \cite{Mi2}.
In particular, it is not obvious how to construct a consistent Hamiltonian
formulation for a particle in a scalar potential in a covariant way.
However, as mentioned above, the classical Snyder dynamics can be formulated only
in Hamiltonian form. To overcome this difficulty,
in this paper we adopt a recently proposed formalism \cite{Mi2} for the coupling
of a particle to a scalar potential that permits the definition of a
covariant Hamiltonian dynamics and the use of the Dirac procedure to eliminate
the constraints. As it is well known, Dirac's
method requires the choice of a time variable, analogous to a gauge fixing,
in order to eliminate the freedom in reparametrization. An important result of our
paper is
that in the interacting Snyder case a consistent choice must necessarily depend on
the dynamics
of the model.

\section{Free particle}
For simplicity we consider the motion in a (1+1)-dimensional spacetime.
Since the structure of classical Snyder spacetime is expressed in terms of its
noncanonical symplectic structure, its dynamics must be written
in Hamiltonian form. For relativistic models, Hamiltonian dynamics can be
described in a covariant way by using the Dirac formalism for constrained
systems \cite{RT,Mu}.

The Snyder fundamental \pb are defined as\footnote{We use units in which $c=1$
and metric signature $(+,-)$.}
\be\lb{PB}
\{x_\m,p_\n\}=\y_\mn+\b p_\m p_\n,\qquad\{x_\m,x_\n\}=\b J_\mn,\qquad\{p_\m,p_\n\}=0,
\ee
where $\y_\mn$ is the flat metric and $J_\mn=x_\m p_\n-p_\m x_\n$ is the generator
of the Lorentz transformations. The parameter $\b$ has dimensions of inverse mass
square and is usually assumed to be of Planck scale.
It can be either positive or negative. In the latter case, the allowed value of the
mass are bounded, $m^2<|\b|^\mo$, as in doubly special relativity \cite{dsr}.
The \pb behave covariantly under Lorentz
transformations, while the action of the translations, generated by $p_\m$, on
spacetime \coo is nonlinear \cite{BK,SM}.

The dynamics of a free particle in Snyder spacetime is trivial.
In fact, since the Lorentz invariance is preserved, the Hamiltonian can be chosen
as in special relativity,
\be
H={\l\over2}(p^2-m^2),
\ee
with $p^2=p_0^2-p_1^2$ and $\l$ a Lagrange multiplier enforcing the mass shell
constraint $\c_1=p^2-m^2=0$.
The Hamilton equations that follow from the nontrivial symplectic structure are
\be
\dot x_\m=\{x_\m,H\}=\l(1+\b p^2)p_\m=\l(1+\b m^2)p_\m,\qquad\dot p_\m=\{p_\m,H\}=0,
\ee
where a dot denotes the derivative with respect to the evolution parameter. 
The constraint $\c_1=0$ is first class, and
according to Dirac, one must impose a further constraint to eliminate the redundant
degrees of freedom $x_0$ and $p_0$ and reduce the system to the motion in one
spatial dimension with external time.

For the standard choice $\c_2=x_0-t=0$, which corresponds to
the identification of the evolution parameter with the coordinate time, one has
\be
C\id\{\c_2,\c_1\}=(1+\b m^2)p_0.
\ee
It follows from the requirement $\dot\c_2=0$ that $\l=1/C$ \cite{Mu}.
On the constraint surface, the dynamics is dictated by the Dirac brackets,
defined as
$$\{A,B\}^*=\{A,B\}+\{A,\c_2\}\,C^\mo\{\c_1,B\}-\{A,\c_1\}\,C^\mo\{\c_2,B\}$$
For the independent variables $x_1$, $p_1$, they read
\be\lb{dirac}
\D\id\{x_1,p_1\}^*=-1,
\ee
as in special relativity.
Moreover, the reduced Hamiltonian $K$ results in
\be\lb{redham}
K=p_0=\sqrt{p_1^2+m^2},
\ee
and the Hamilton equations following from (\ref{dirac}) and (\ref{redham}) are
\be\lb{eqfree}
{dx_1\over dt}={p_1\over\sqrt{p_1^2+m^2}},\qquad{dp_1\over dt}=0,
\ee
which coincide with the \eom of a free particle in special
relativity. In the case of a free particle the motion in Snyder spacetime is
therefore trivial.

In view of the the results on the interacting particle of the following section,
it appears however that a more physical choice of gauge is given by a constant
rescaling of time, $t=\sqrt{1+\b m^2}\ x_0=\sqrt{1+\b p^2}\ x_0$. A justification
for this choice is that the natural metric of spacetime, invariant under Snyder
transformations is $ds^2=(1+\b p^2)dx^2$, with $dx^2$ the Minkowski metric \cite{SM}.

In this gauge, $\{\c_2,\c_1\}=(1+\b m^2)^{3/2}p_0$, but the \db are still given by
(\ref{dirac}). The reduced Hamiltonian is now
\be
K={p_0\over\sqrt{1+\b m^2}}=\sqrt{p_1^2+m^2\over1+\b m^2},
\ee
and the Hamilton equations read
\be
{dx_1\over dt}={p_1\over\sqrt{(1+\b m^2)(p_1^2+m^2)}},\qquad{dp_1\over dt}=0.
\ee
Of course, the only difference from (\ref{eqfree}) is a rescaling of the momenta.
In this gauge the rest energy of a particle is $m_0=m/\sqrt{1+\b m^2}$.

\section{Harmonic oscillator}
A more interesting problem occurs when the particle is subject to an external force
generated by a potential.
We shall consider in particular the case of a harmonic potential, which depends only
on the spatial position of the particle, $V=V(x_1)$. To our knowledge, the coupling
of a particle with a scalar potential in classical special relativity has not been
discussed in depth.
Here, we adopt the proposal of \cite{Mi2}, that preserves the reparametrization
invariance of the theory and hence permits the use of the Dirac formalism.
According to it, a consistent Hamiltonian for a particle coupled to a scalar potential
is given by
\be\lb{ham}
H={\l\over2}[p^2-(m+V)^2]=0,
\ee
enforcing the constraint $\c_1=p^2-(m+V)^2=0$.
The \eom derived from (\ref{ham}) with the help of the \pb (\ref{PB}) read
\bea\lb{hameq}
&\dot x_0=\l[(1+\b p^2)p_0+\b J(m+V)V'],\qquad&\dot p_0=\l p_0p_1(m+V)V',\cr
&\dot x_1=\l(1+\b p^2)p_1,\qquad&\dot p_1=-\l(1-\b p_1^2)(m+V)V',
\eea
where a prime denotes a derivative with respect to $x_1$ and $J\id J_{10}$ is
the generator of the Lorentz transformations.
Note in particular that, because of the nontrivial \pb between $x_0$ and $x_1$, an
additional term proportional to $\b$ appears in the $\dot x_0$ equation in
comparison with special relativity.
Moreover, due to the nontrivial symplectic structure, in the limit $c\to\inf$, with
$\b$ constant, the coordinate $x_0$ does not coincide with the nonrelativistic time,
and hence in that limit the relativistic Snyder dynamics does not go into the
nonrelativistic Snyder dynamics.

In the interacting case, it is difficult to find a gauge fixing compatible with the
nontrivial symplectic structure. For example, the gauge choice $t=x_0$ leads to
inconsistencies. One is forced to make a choice of time that depends on the dynamics
of the model. We choose
\be
\c=Sx_0-t=0,
\ee
with
\be
S=\sqrt{1+\b(m+V)^2}=\sqrt{1+\b p^2}.
\ee
As explained before, this choice can be understood considering the natural metric of
the Snyder spacetime \cite{SM}. This choice will make the Dirac brackets independent
of $x_0$.

The Poisson bracket of the constraints $\c_1$ and $\c_2$ reads
\bd
C=\{\c_2,\c_1\}=S(1+\b p^2)p_0+\b SJ(m+V)V'+\b S^\mo(1+\b p^2)J(m+V)V'p_1x_0
\ed
\be
= S[S^2+(m+V)V'x_1]p_0.
\ee
The Lagrange multiplier resulting from this gauge choice is therefore $\l=1/C$,
and the \db of the independent variables $x_1$ and $p_1$ read
\be\lb{intdirac}
\{x_1,p_1\}^*=-{S^2\over S^2+(m+V)V'x_1}.
\ee

The reduced Hamiltonian $K$ must be chosen so that it generates the motion on the
reduced phase space induced by the \db (\ref{intdirac}). The correct choice is
\be\lb{intham}
K={p_0\over S}=\sqrt{p_1^2+(m+V)^2\over1+\b(m+V)^2}.
\ee
This quantity is conserved under the time evolution dictated by $H$ and represents
the energy of the system in the laboratory frame.

The Hamilton equations derived from (\ref{intdirac}) and (\ref{intham}) or
equivalently from (\ref{hameq}) are
\be\lb{redhameq}
{dx_1\over dt}={1\over B}\,{p_1\over K},\qquad
{dp_1\over dt}=-{1-\b p_1^2\over(1-\b m^2)B}\,{(m+V)V'\over K},
\ee
where $B=S^2+(m+V)V'x_1$.

This system of equations can be solved in two steps. First define an
auxiliary time variable $\t$, such that $dt=B d\t$, so that the equations
(\ref{redhameq}) take the form
\be\lb{rehameq}
{dx_1\over d\t}={p_1\over K},\qquad
{dp_1\over d\t}=-{1-\b p_1^2\over1-\b m^2}\,{(m+V)V'\over K}.
\ee
The equations (\ref{rehameq}) can be solved in a standard way by exploiting the
conservation of the reduced Hamiltonian $K$, from which follows
\be
p_1^2=K^2-(1-\b K^2)(m+V)^2
\ee
and then, using the first of eqs.\ (\ref{rehameq})
\be\lb{oscill}
{dx_1\over d\t}=\sqrt{1-\left({1-\b K^2\over K^2}\right)(m+V)^2}.
\ee

A redefinition of the energy,
\be
E={K\over\sqrt{1-\b K^2}},
\ee
reduces this equation to that of classical special relativity. In particular,
in the case of the harmonic
oscillator with potential $V={\k\over2}x_1^2$, the solution is \cite{Mi2,Ha}
\be\lb{pos}
x_1=\sqrt{E^2-m^2\over\k E}\ \sd(\o\t,q),
\ee
where
$$\o^2={\k\over E}={\k\sqrt{1-\b K^2}\over K},\qquad q={E-m\over2E}=
{K-m\sqrt{1-\b K^2}\over2K},$$
and $\sd(\o\t,q)$ is a Jacobian elliptic function.
The period of oscillation
$T_0$ can be written in terms of the complete elliptic integral {\bf K}$(q)$ as
\be
T_0={4\over\o}\ {\rm\bf K}(q)\sim{2\p\over\o_0}\left(1-{3\over8}\,{E-m\over m}\right),
\ee
where $\o_0=\sqrt{\k/m}$ is the frequency of the nonrelativistic oscillator and we have
written the first terms of a low-energy expansion.

After some calculations, one gets for the momentum
\be\lb{mom}
p_1=\sqrt{E^2-m^2\over1+\b E^2}\ \cd(\o\t,q)\nd(\o\t,q),
\ee
with $\cd(\o\t,q)$ and $\nd(\o\t,q)$ Jacobian elliptic functions.

One can now write down the solution of the Snyder oscillator in terms of the physical
time variable $t$, substituting (\ref{pos}) and (\ref{mom}) in its definition and
integrating,
\bea\lb{time}
t&=&\int[1+\b(m+V)[(m+V)+V'x_1]d\t=\int\left[m^2+2m\k x_1^2+{3\over4}\,\k^2x_1^4\right]
\,d\t\cr
&=&(1+\b E^2)\t-{\b(E^2-m^2)\over\o}\,\sd(\o\t,q)\cd(\o\t,q)\nd(\o\t,q).
\eea
Eqns.\ (\ref{pos}), (\ref{mom}) and (\ref{time}) give the exact solution of the
relativistic Snyder oscillator in parametric form.
The period of oscillation is given by $T=t(T_0)$, i.e.
\be\lb{period}
T={4\,(1+\b E^2)\over\o}\ {\rm\bf K}(q)={4\over(1-\b K^2)\,\o}\ {\rm\bf K}(q),
\ee
and contains energy-dependent contributions coming from both special relativity and
Snyder dynamics.

It may also be interesting to consider the limit of the solution for small energy, namely
$K-m_0\ll m\ll\b^{-1/2}$. In this limit one can make an expansion in powers of
${K-m_0\over m_0}$. However, to keep the formulae simple we shall write the expansion
in terms of $E$ and $m$. It is easy to check
that  ${E-m\over m}\sim {K-m_0\over m_0}\,(1+\b m_0^2)$.

At first order in ${E-m\over m}$, the corrections can be obtained from the
nonrelativistic limit of eqns.\ (\ref{pos}) and (\ref{mom}),
\be
x_1\sim\sqrt{2(E-m)\over\k}\ \sin\o_0\t,\qquad p_1\sim\sqrt{2m(E-m)\over1+\b m^2}\
\cos\o_0\t,
\ee
with
\be
t\sim(1+\b m^2)\t-{\b m(E-m)\over\o_0}\,\sin\t\cos\t,
\ee
and the period $T$ of the oscillator is given by
\be\lb{period}
T\sim{2\p\over\o_0}\left[1+\b m^2-\left({3\over8}-2\b m^2\right){E-m\over m}\right].
\ee
For positive $\b$ the period is increased with respect to special relativity.
It may be compared with the exact nonrelativistic result \cite{Mi},
$T={2\p\over\o_0}\,[1-2\b m(E-m)]^{-1/2}\sim{2\p\over\o_0}\,[1+\b m(E-m)]$.
The limit of (\ref{period}) for $c\to\inf$, with $\b$ fixed, does not coincide with this
result. The reason is of course that the time variable of the relativistic
model differs from that of the nonrelativistic model in this limit.

\section{Conclusions}
We have investigated the dynamics of the harmonic oscillator for the relativistic Snyder
model and found an exact analytic solution of the equations of motion.
The solution presents corrections of order $\b E^2$ with respect to that of special
relativity, as could have been predicted from dimensional considerations.
In particular, for positive $\b$, the period of the harmonic oscillator is increased with
respect to that of the relativistic oscillator.

An interesting result of our investigation is that in the interacting case the choice
of the time variable must depend on the dynamics. This can be understood considering
that in Snyder spacetime the natural metric is given by $(1+\b p^2)(dx_0^2-dx_1^2$), and
hence time in the laboratory frame is measured by $t=\sqrt{1+\b p^2} x_0$.

As for the nonrelativistic case \cite{Mi3}, our results can easily be generalized to a
curved background. A more challenging problem would be to extend our investigation to the
quantum dynamics. This is of course not straightforward, since it would be necessary to
define a quantum field theory framework.

\section*{Acknowledgements}
I wish to thank  Stjepan Meljanac and Boris Iveti\'c for useful comments.

\end{document}